# Apparent magnetic field decay in a solar flare from Zeeman splitting of the Mg I 12.32 μm observations

Yuanyong Deng[1,2,*], Xianyong Bai[1,2,3,*], Jie Hong[4], Wenxian Li[1,2], Jinliang Wang[1,5], Yuyang Ye[1,2], Kaifan Ji[6], Jiangtao Su[1,2], Yongliang Song [1], Yuliang Shen[1], Ziyao Hu[1,2], Xiao Yang[1], Dongguang Wang[1], Jiaben Lin[1], Junfeng Hou[1,2], Xiaoming Zhu[1], Xing Hu[1], Zhiwei Feng[1], Wei Duan[1], Biyuan Gao[1], Liyue Tong[1],Yang Bai[1], Jianwen Hua[7], Lianwei Zhao[7], Songbo Xu[8], Shoulin Wei[9]

1.State Key Laboratory of Solar Activity and Space Weather, National Astronomical Observatories, Chinese Academy of Sciences, Beijing, China; 2. School of Astronomy and Space Science, University of Chinese Academy of Sciences, Beijing, China; 3. Institute for Frontiers in Astronomy and Astrophysics, Beijing Normal University, Beijing, China; 4. Institute for Solar Physics, Department of Astronomy, Stockholm University, Stockholm, Sweden; 5. School of Computational Science and Electronics, Hunan Institute of Engineering, Xiangtan, China; 6. Yunnan Astronomical Observatory, Chinese Academy of Sciences, Kunming, China; 7. Shanghai Institute of Technical Physics, Chinese Academy of Sciences, Shanghai, China; 8. Xi'an Institute of Optics and Precision Mechanics, Chinese Academy of Sciences, Xi'an, China; 9. Faculty of Information Engineering and Automation, Kunming University of Science and Technology, Kunming, China; e-mail: dyy@nao.cas.cn; xybai@nao.cas.cn;

The magnetic field is the most fundamental observable in solar physics, while solar flares are the most violent, rapid magnetic energy-release events in the solar atmosphere. Decades of observations and theory have sketched the framework of flare evolution and magnetic-field-related physical process, yet the actual magnetic field variations during flares remained poorly understood. The main obstacle is that magnetic-diagnostic spectral lines are distorted by transient flare activity, making traditional spectro-polarimetry inversions complicated. Using a mid-infrared solar telescope, we investigate spectroscopic evolution of the Mg I 12.32 μm line during an X1.5-class flare. Owing to the exceptionally large Zeeman sensitivity of this line, magnetic field strength can be directly determined from the measured line separation. Our analysis reveals apparent magnetic field weakening beneath flare ribbons in the impulsive phase. Radiative-hydrodynamic simulations and atmospheric inversions confirm that the formation height of the Mg I 12.32 μm line increases during the flare. We suggest this apparent field change probably originates either from elevated line-formation altitude or magnetic-field redistribution. Additional observations and modeling are needed to clarify these two contributions. This study demonstrates the capability of mid-infrared wavelengths to trace magnetic-field evolution during flares.

## Introduction

Solar flares are explosive energy-dissipation events driven by magnetic reconnection[1] on timescales as short as minutes, triggering catastrophic space-weather events in the solar system. According to the standard flare model[2], coronal magnetic fields undergo rapid topological rearrangement: a magnetic flux rope erupts upwards away from the reconnection site, while post-flare loops form below this site. Although major energy dissipation take place at coronal altitudes, observational findings indicate that the lower-lying photosphere is also disturbed[3]. The horizontal component of photospheric magnetic field has an abrupt, permanent increase near the flaring polarity inversion line (PIL)[4-7]. Rapid sunspot or pores rotation are also found during large flares[8-11]. These results indicate immediate feedback from coronal eruptions to the photosphere[12].

While measurements of magnetic fields before and after flares are robust, characterizing their evolution within the lower solar atmosphere during flares remains exceptionally challenging. Spectral line shapes change dramatically over short periods as flares erupt[13-16], demanding high-cadence observations and non-local thermodynamic equilibrium (NLTE) atmospheric models to properly invert the highly dynamic flare atmosphere. High-cadence spectropolarimetry with photospheric and chromospheric lines from ground-based facilities, including the Swedish 1-m Solar Telescope[17], the 1.5-m Gregor solar telescope[18], the 1.6m Goode Solar Telescope[19], and the 4-m Daniel K. Inouye Solar Telescope[20], now enables such measurements. Meanwhile, advanced NLTE inversion codes, such as the Stockholm inversion code (STiC)[21], NICOLE[22], and DeSIRe[23], have become available recently. These state-of-the-art observations and techniques have uncovered magnetic field variations within flare ribbons across C, M, and X-class events[24-29]. However, because linear polarization signals are typically weak, only the line-of-sight (LOS) component of the magnetic field can usually be retrieved, especially for chromospheric lines. This raises a key question: can changes of magnetic field strength other than the LOS component during flares be measured accurately during flaring activity?

The Zeeman effect is the foundation of solar magnetic field measurements[30], and all aforementioned studies rely on this effect. According to the Zeeman splitting formula, the spectral line splitting is proportional to the square of the wavelength λ and the Landé factor g. The thermal line width is proportional to the wavelength. In other words, the Zeeman sensitivity (gλ) becomes far more pronounced at mid-infrared wavelengths[31], presenting a unique opportunity to determine magnetic field strengths by direct measurement of the splitting. This model-independent method provides an alternative way to study the evolution of magnetic field during flares.

## Observations and results

"Accurate Infrared Magnetic field Measurements of the Sun" (AIMS) is a recent developed mid-infrared solar telescope [32], equipped with a imaging Fourier Transform Infrared Spectrometer (FTIR) that uses a 128-pixel (64×2) mercury cadmium telluride (HgCdTe) detector. For each pixel, an interferograms containing 3.226 million interference points is acquired within 30 s under a maximum optical path difference of 132 cm, yielding a pixel sampling rate of 107.53 kHz. Its inverted spectrum of the target Mg I 12.32 μm has a spectral resolution of about 0.06 nm, corresponding to a pixel velocity resolution of 1.46 km/s. The field of view (FOV) is 96"x3" with a pixel scale of 1.5". AIMS can measure magnetic field strength directly using Zeeman splitting measurements for the solar regions with field strengths greater than 300 G. A general formation height of the Mg I 12.32 μm line core is in the upper photosphere, about 450 km for quiet-sun region[33]. AIMS is also equipped with an 8-10 μm wide-band imaging terminal (ITS), which uses a 256×256 HgCdTe detector with a 1 s cadence and pixel scale of 1.5". With a 1-meter aperture, the diffraction-limited spatial resolution of AIMS at 12.32 μm is around 3″.

We investigate an X1.5 flare that occurred on 30 December 2024 in NOAA Active Region 13936 (10° N, 46° W; Fig. 1). Fig. 1a presents the 1600 Å image formed at the temperature minimum region captured near the hard X-rays emission peak. Distinct, elongated flare ribbons are clearly visible. Flare-released energy propagated downward through the chromospheric and photospheric layers, triggering the white-light flares in both the mid-infrared continuum at 8-10 μm and the visible images at 617.3 nm, as displayed in Fig. 1b and 1c (also seen Supplementary Video 1). The flare ribbons are located near the magnetic PIL from the radial magnetogram in Fig. 1d.

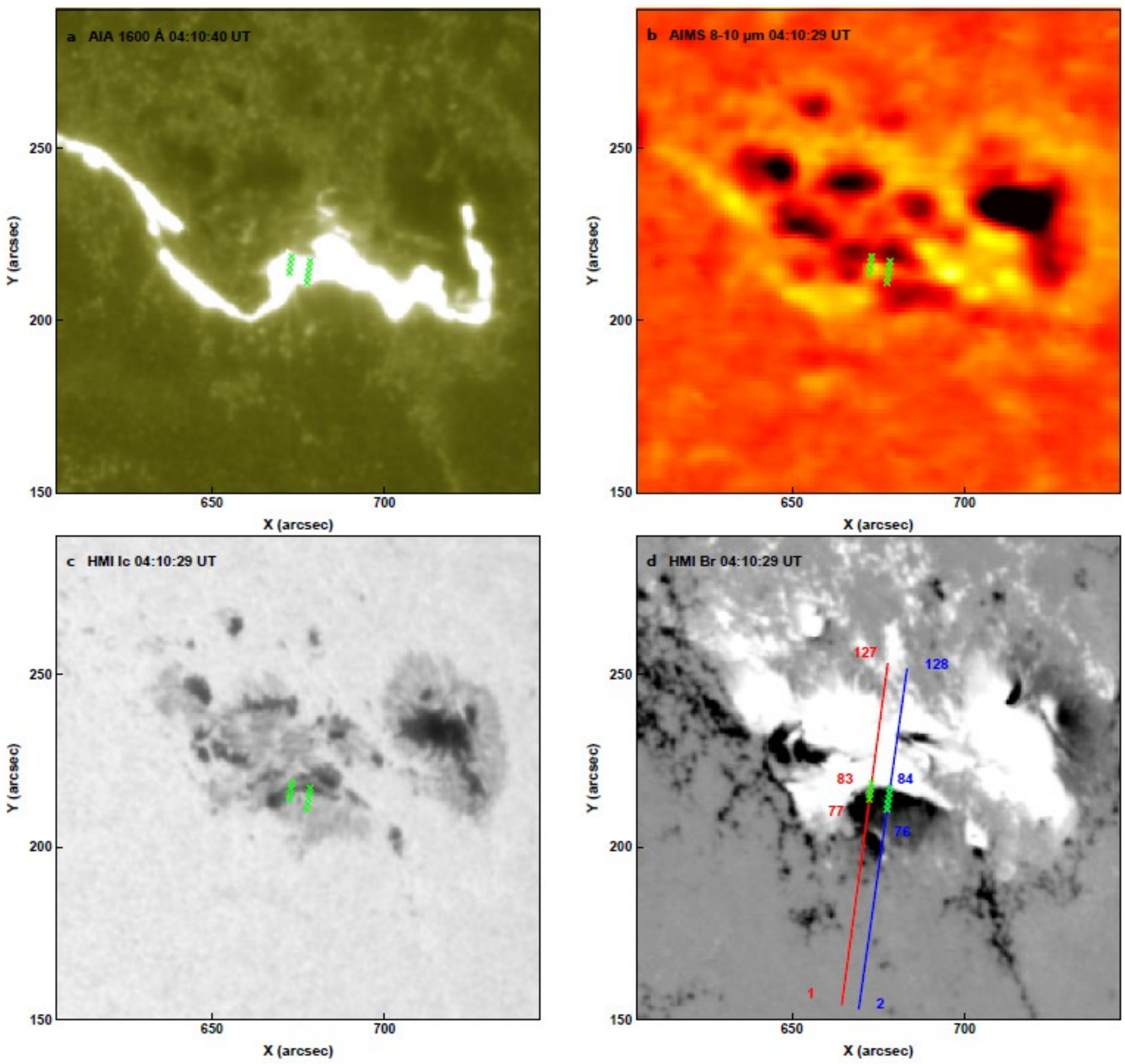

Fig. 1 | Multiwavelength observations of the X1.5-class flare on 30 December 2024. a, EUV 1600 Å image from the Atmospheric Imaging Assembly (AIA) on board the Solar Dynamics Observatory (SDO) [34], showing the flare ribbons. b, Mid-infrared 8–10 μm image obtained by AIMS. c and d: Photospheric continuum image and radial magnetogram from the Helioseismic and Magnetic Imager (HMI) on board SDO. All panels use the same helioprojective field of view; X and Y are given in arcseconds, and

coordinate tick labels are shown in every panel. Red and blue lines in panel d mark the odd- and even-numbered pixels of the AIMS FTIR 64×2-pixels, respectively. The two 64×1 linear arrays are separated by approximately 1.5”. The observations were obtained in sit-and-stare mode. The nine green crosses mark pixels 76–84, in which Mg I 12.32 μm line emission is detected during the flare and which are analyzed in detail. The corresponding temporal evolution is shown in Supplementary Video 1.

The flare started at 04:01 UT, reached its peak in GOES soft X-rays (SXR) at 04:14 UT, and ended at 04:38 UT, with a total duration of 37 minutes, as shown in Fig. 2a. The hard X-ray peak of the flare, acquired by the Hard X-ray Imager (HXI) on board the Advanced Space-based Solar Observatory (ASO-S)[35], occurred at ~04:10 UT, which is four minutes before the SXR peak. The orientation of the AIMS FTIR detector array is perpendicular to the PIL, and we obtained continuous data across all flare phases, from 04:01 to 04:38 UT. Fig. 2b and 2c show the Mg I 12.32 μm spectrum before and near the flare maximum, respectively. Both the mid-infrared continuum and the Mg I 12.32 μm line for pixels 76-84 (marked by arrows) were enhanced during the flare. These pixels correspond to the flare ribbon regions, in the penumbra and umbra, as indicated by the green crosshairs in Fig. 1b and 1c. Furthermore, well-resolved triple Zeeman splitting profiles are observed in these pixels. We applied a three-component Gaussian fit to these profiles using the Markov Chain Monte Carlo (MCMC) method **(**methods and Extended Data Fig. 1**)** to derive the magnetic field strength. An example of the observed and fitted profiles for pixel 82 at four times during the flare is shown in Fig. 2d.

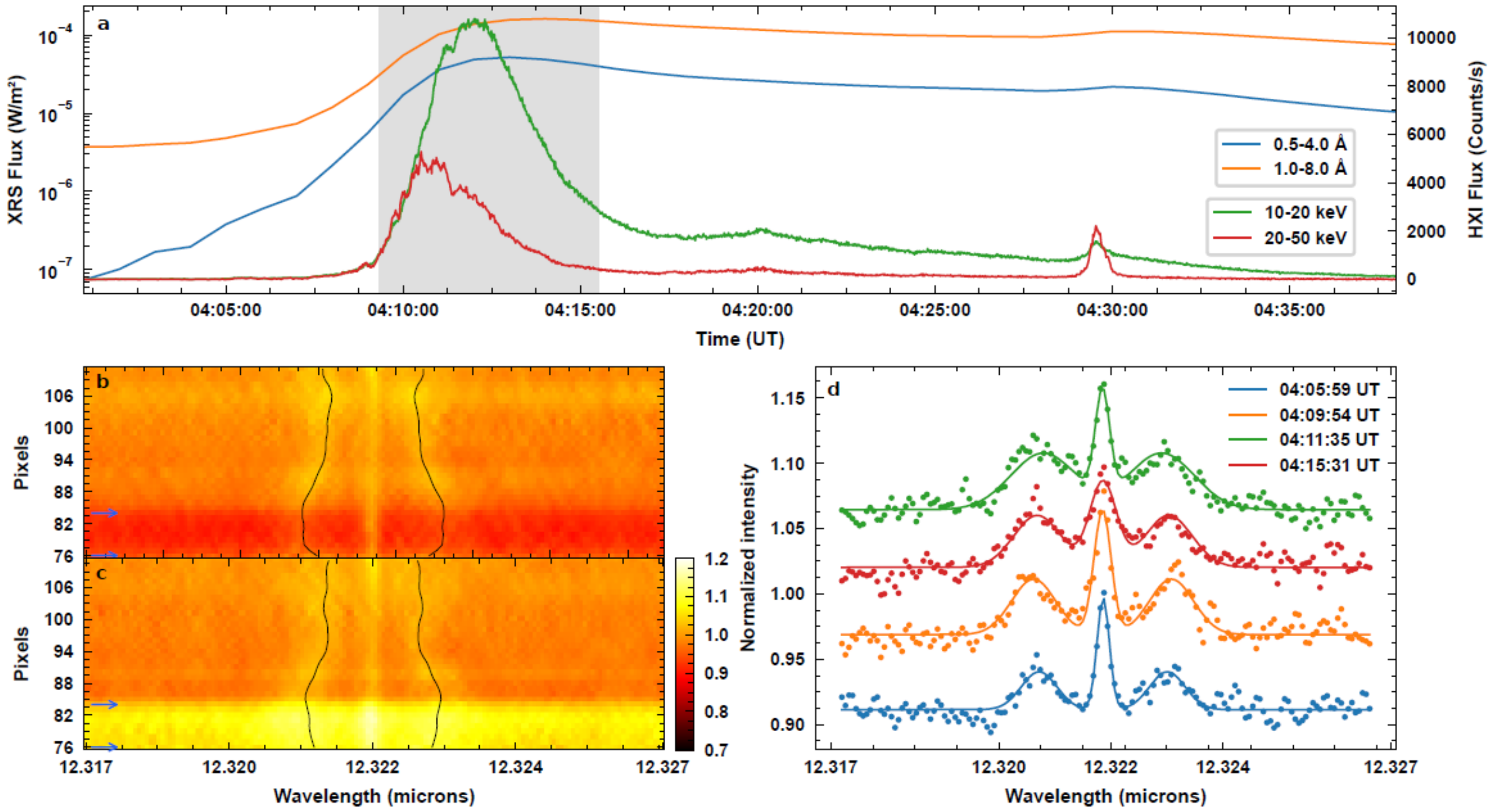

Fig. 2 | X-ray light curves and Mg I 12.32 μm spectral evolution of the X1.5 flare. a, X-ray light curves: blue and orange show GOES soft X-ray fluxes at 0.5–4 Å and 1–8 Å, respectively; green and red show ASO-S/HXI count rates at 10–20 keV and 20–50 keV, respectively. The grey interval marks 04:09:20–04:15:31 UT, during which the magnetic-field decrease is identified. b and c, Mg I 12.32 μm spectra for pixels 76–111 before the flare (b, 04:05:59 UT) and near the flare maximum (c, 04:11:35 UT). The arrows mark pixels 76–84, where the continuum and line emission are enhanced. The two thin black curves trace the fitted positions of the two σ-component peaks at each pixel. d, Mg I 12.32 μm profiles for pixel 82 at 04:05:59, 04:09:54, 04:11:35 and 04:15:31 UT (blue, orange, green and red, respectively). The points are the processed observations and the solid curves are the posterior-median three-component Gaussian profiles from the MCMC fit.

Fig. 3 shows the temporal evolution of the continuum intensity and magnetic field strength for pixels 76-111 from 04:01 to 04:38 UT, covering the entire flare phase. The continuum intensity is normalized to the median value of the pre-flare times from 04:01 to 04:06 UT for each pixel. A clear decay in magnetic field strength is found for all nine analyzed pixels 76-84, as shown in Fig.3b. This decay began at 04:09 UT, near the hard X-ray flare peak, and ended at 04:15 UT, which is about 90 s after the SXR, meaning the primary magnetic field change occurred during the flare's impulsive phase. In Fig. 3c, the magnetic field strength decays from 1593.92 to 1474.8 G, yielding a field reduction of 119.12 G for pixel 78. By comparison, Fig. 3d shows a magnetic field decrease from 1644.79 G to 1533.66 G for pixel 82, corresponding to a decay magnitude of 111.13 G. For the other sampled pixels, the magnetic field varies spatially from 1395.87 G to 1665.21 G (Extended Data Fig. 2 and Extended Data Table 1). Across the nine analyzed pixels, the measured field decay magnitudes range from 53.89 G to 148.4 G (Extended Data Table 1). A statistical significance test of this magnetic field decay is also carried out, as shown in Extended Data Table 1. The mid-infrared continuum intensity increased by ～15% (with slight variations among pixels), peaking at 04:13:20-04:14:20 UT, coinciding with the SXR flare peak. The magnetic fields exhibited a recovery tendency after 04:18 UT.

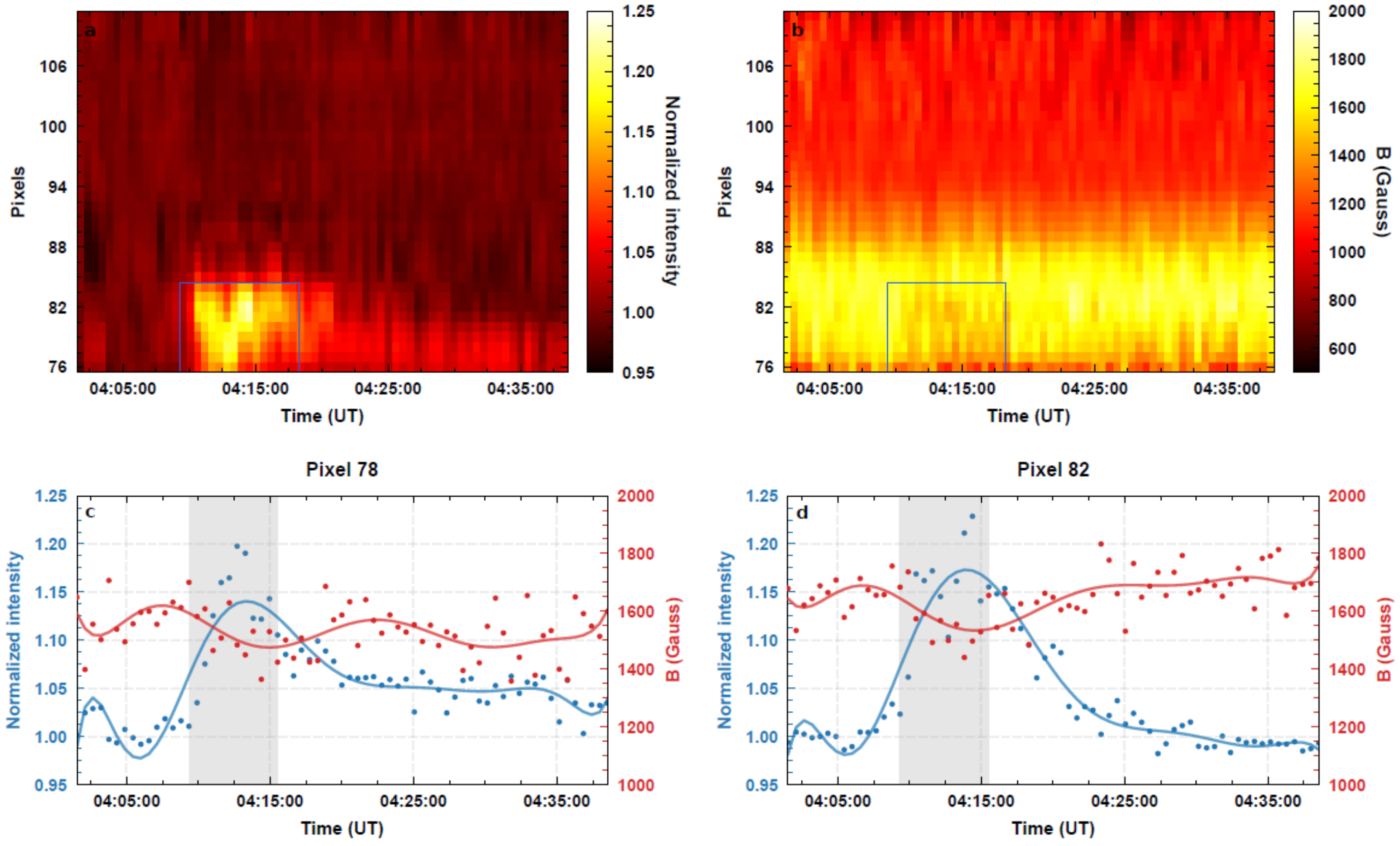

Fig. 3 | Spatiotemporal evolution of the mid-infrared continuum intensity and magnetic-field strength. a,b, Time–pixel maps of the normalized continuum intensity (a) and magnetic-field strength (b) for AIMS FTIR detector pixels 76–111. The blue boxes delimit pixels 76–84 and the 04:09:20–04:18:19 UT flare-affected interval used separately in the background-trend analysis. c,d, Temporal evolution of the normalized continuum intensity (blue points) and magnetic-field strength (red points) for pixels 78 (c) and 82 (d); the corresponding solid curves are descriptive ninth-degree polynomial fits. The grey shading in c and d marks the magnetic-field-decrease interval 04:09:20–04:15:31 UT. The longer 04:09:20–04:18:19 UT interval masked in the background-trend analysis additionally includes the early recovery and is not the interval represented by the grey shading.

## Simulations

To verify the observed decay, we also performed radiative-hydrodynamic simulations and spectral inversion. The Mg I 12.32 μm line response during a flare is modelled using the 1D radiative-hydrodynamic code RADYN and spectral synthesis code RH1.5D (Methods) [36]. We take the same flare model and recalculate the polarized line profiles adding a prescribed time-independent magnetic field. The magnetic field strength varies with height as an exponential distribution of $B(z)=B_0 \exp\left(-\frac{z}{H}\right)$. The values $B_0$ and H are set to 1765.75 G and 1741.46 km, respectively. The inclination and the azimuth of the magnetic field are fixed to 45 degrees.

The simulation shows that both the mid-infrared continuum and the Mg I 12.32 μm line core intensity are enhanced (Fig. 4a) after a certain time, due to an increased sensitivity to the local temperature and heated atmosphere. Our observation results (Fig. 2d) are consistent with the simulations. The line's formation height increases as heated chromospheric layers contribute more to line emission during flares. We calculated the magnetic field weighted by the response function of the Zeeman splitting (the solid line in Fig. 4b), which closely agrees with the magnetic field inferred directly from the synthetic Zeeman splitting profiles (the dashed line in Fig. 4b) in both magnitude and temporal evolution. Both quantities show a pronounced decrease during the initial phase and remain below their initial values throughout the simulation. The close agreement therefore supports the use of the Zeeman splitting as a diagnostic of the evolution of the effective magnetic field sampled by the Mg I line.

To further diagnose the temperature variation during the flare, we performed inversions of the observed Mg I 12.32 μm line profile using the STiC[21] code (Extended Data Fig. 3 and Methods). In the inversion, we utilized the model atom Mg I_66.atom[37] and adopted the FAL-C model atmosphere[38] as the initial guess. Since our purpose lies in the temperature variations during the flare, we employ five nodes for temperature, while using only one node for both the magnetic field and velocity. We performed inversions for spectra taken from the flare ribbon region for pixel 82, obtaining the temporal evolution of the temperature stratification.

Fig. 4c presents the inverted temperature stratification as a function of height for pixel 82 at three different times. The plot reveals that at 04:11 UT, following the flare onset, the height of the temperature-minimum region decreases, which is consistent with previous flare dynamic models[14-15]. Additionally, elevated temperatures are observed in both the photosphere and chromosphere. This provides a physical explanation for the observed enhancement of mid-infrared continuum and Mg I 12.32 μm line core emission. The normalized response function of the Mg I 12.32 μm line core to changes in temperature (Fig. 4d) clearly demonstrates an increased chromospheric contribution during the flare, consistent with the simulation[36].

Combining RADYN radiative-hydrodynamic simulations and STiC inversions of the observed spectra, we propose a physical scenario: the formation height of the Mg I 12.32 μm line core shifts to higher chromospheric layers during solar flares, and the intrinsically weaker magnetic fields at these altitudes give rise to the measured decay in the magnetic field strength.

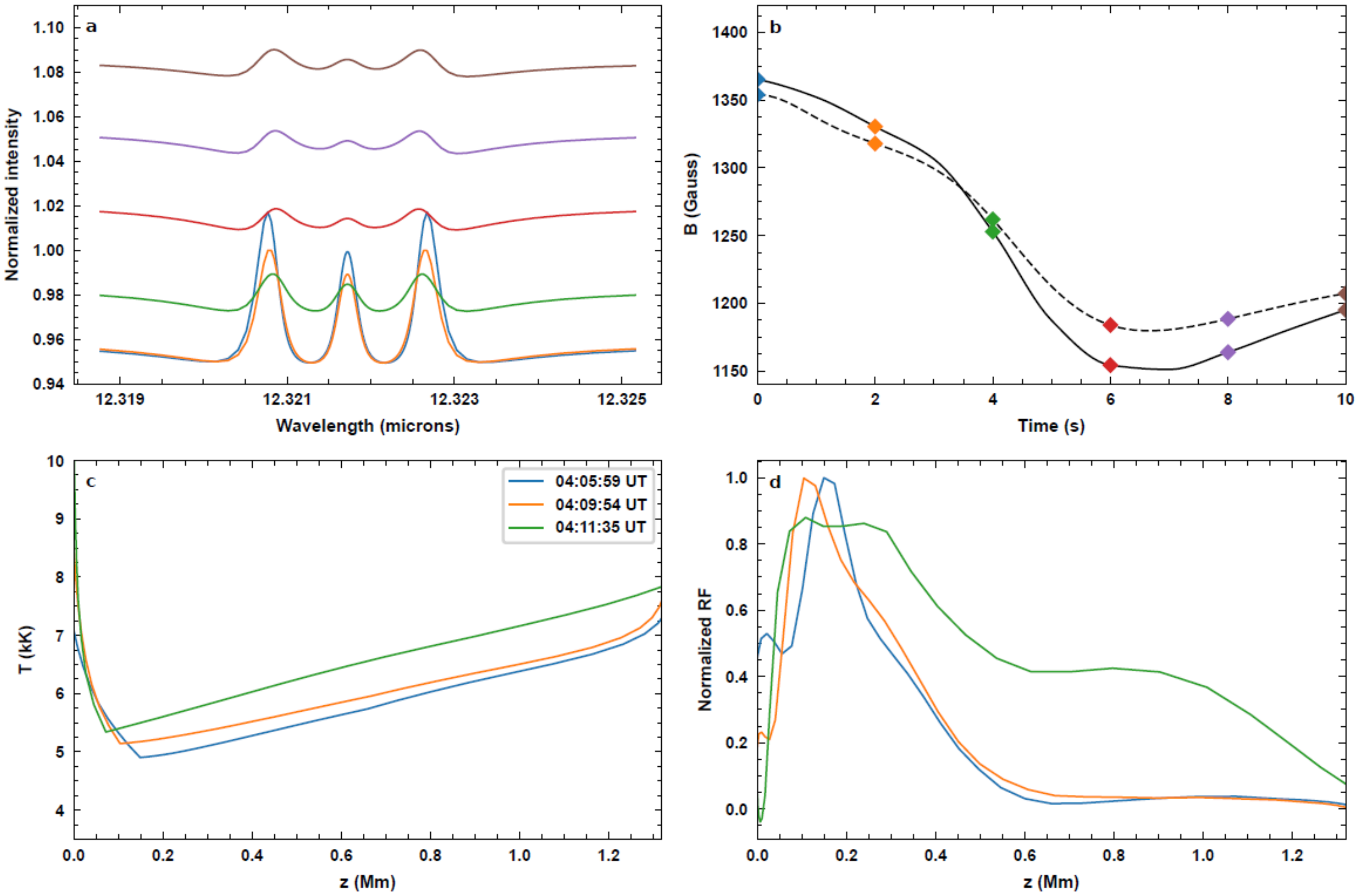

Fig. 4 | Radiative-hydrodynamic simulations and spectral inversion. a, Synthetic Mg I 12.32 μm spectra from the RADYN model and RH synthesis at simulation times 0, 2, 4, 6, 8 and 10 s, shown in blue, orange, green, red, purple and brown, respectively. b,

Weighted magnetic fields strength (solid line) in the RADYN model and measured magnetic field strength (dashed line) from the Zeeman splitting of the synthetic profiles in panel a. Different colors of diamonds correspond to different simulation times. c, Inferred temperature stratification with height from STiC inversions of the observed spectra for pixel 82 (blue:04:05:59, orange:04:09:54, green:04:11:35 UT). d, Normalized response function of the Mg I 12.32 μm line core to changes in the temperature.

## Conclusion and Discussion

Using a 128-pixel FTIR on the AIMS telescope, we have obtained complete spectroscopic evolution of the triple Zeeman splitting Mg I 12.32 μm line during an X1.5 flare. Our analysis reveals an apparent weakening of the magnetic field strength beneath flare ribbons during the flare impulsive phase, with a subsequent recovery tendency. Radiative-hydrodynamic simulation and spectral inversions confirm that this change is probably a consequence of the Mg I 12.32 μm line forming at elevated chromospheric altitudes during the flare. We used simulated flare spectra to replicate the measurement process, which also led to the conclusion that the magnetic field weakens during the flare. As the magnetic field topology is changed due to the magnetic reconnection[39], the magnetic field distribution with height also changes during the flare, which is not considered by our simulation. It is inevitable that our observed decay of magnetic field strength is coupled with the magnetic field evolution during the flare.

As the flare enters the decay phase, the solar atmosphere undergoes gradual cooling, which lowers the formation height of the Mg I 12.32 μm line. Consequently, the altitude corresponding to the measured magnetic field drops, leading to the recovery in the observed field strength. Notably, newly reconnected magnetic field lines produced during the flare above the PIL contract downward under magnetic tension and eventually settle in the photosphere or lower chromosphere under buoyancy equilibrium. These field lines also facilitate the recovery of the magnetic field strength measured by the Mg I line. However, current observations are insufficient to quantify the relative contributions of these two scenarios. Additional measurements and modeling are needed to clarify these two contributions.

The continuous observations of the Mg I 12.32 μm line after the peak of an X5.7-class flare (or the decay phase) is obtained using a single-pixel detector in the 1990s[40], reporting a sustained magnetic field weakening. However, the positional uncertainties due to the limb guiding errors may have affected the result[40]. Our study focuses on the flare impulsive phase, rather than the decay phase. Moreover, all nine pixels show the decay of magnetic fields and the influence of guiding system (the root-mean-square value of tracking errors is 0.88”) can be ignored. Significant weakening of the longitudinal magnetic field at multiple locations during a major flare is detected using Ni I 676.8 nm line[41]. The magnetic field strength evolution can not be determined due to a lack of transverse components. This line also forms at higher altitudes during flares[13], suggesting their observed longitudinal weakening may also arise from an increase in line formation height, which is consistent with our results.

Numerous studies have reported enhanced photospheric transverse magnetic field along the PIL during flares[4-7], a phenomenon explained by the magnetic implosion model[12]. The magnetic field strength weakening phenomenon we observed is not contradictory with these results. As demonstrated by our simulations, the Mg I 12.32 μm forms at an elevated height corresponding to chromospheric layers during the flare’s impulsive phase. The chromospheric region beneath flare ribbons lies below the magnetic diffusion region, and its magnetic field decays directly due to magnetic reconnection. Our result is consistent with the early observation by Kleint[42], who found that the absolute value of chromospheric line-of-sight magnetic field decreased in about 62% of the observed pixels during an X1-flare. As the reconnected magnetic field falls back to the photosphere, it superimposes and accumulates there, leading to the observed enhancement of the horizontal magnetic field strength.

Then why does the magnetic field accumulate in the photosphere rather than in the chromosphere? A recent study[43] indicates that the magnetic field has to overcome a strong magnetic buoyancy to sink into the photosphere, a process only possible for the magnetic field lines with a spatial scale of ＜10 km. Most magnetic loops do not meet this criterion, so the reconnected magnetic field accumulates in the photosphere.

This work demonstrates the power of mid-infrared Zeeman splitting measurements for probing solar magnetic field evolution during flares, a capability previously unavailable with conventional spectropolarimetric methods. Future observations with AIMS will enable high-precision tracking of magnetic field evolution during flares for more cases and different classes. Coordinated multi-line spectroscopic observations from AIMS, the SST, GREGOR and DKIST, deliver the possibility of inferring the thermal properties and the magnetic field at multiple heights with high accuracy. Such datasets also help to quantitatively investigate the respective contributions of formation-height effects and the intrinsic modifications to the magnetic field distribution driven by magnetic reconnection.

## Methods

### FTIR Data Processing and magnetic field determination

The FTIR data were processed using our interferogram-to-spectrum pipeline. First, a triangular window function was applied to the interferograms to suppress spectral leakage. The windowed interferograms were then transformed into the spectral domain using a fast Fourier transform (FFT), producing the corresponding spectral images for all 128 pixels. As the wavelength range of the inverted spectra is about 0.1 μm, only the spectral segment surrounding the Mg I 12.32 μm line was extracted for subsequent analysis.

The selected spectra were corrected for detector non-uniformity using flat-field calibration. A Gaussian fitting procedure was then applied to obtain the spectral curvature parameters for the 128 pixels. To mitigate residual stripe-like artifacts in the data, two-dimensional median filtering was applied along the horizontal direction to suppress transverse stripe patterns, while wavelet-based filtering was performed in the vertical direction to reduce longitudinal noise components. Finally, a spectral curvature correction was applied to remove residual distortions in the spectrogram.

Magnetic field strengths were derived from quantitative measurements of the triple Zeeman splitting profiles in the Mg I 12.32 μm line. A three-component Gaussian function is employed to fit the observed line profiles, with the following formula:

$$I(\lambda)=a_1+a_2\exp\left(-\frac{(\lambda-a_3)^2}{2a_4^2}\right)+a_5\left[\exp\left(-\frac{(\lambda-a_3-a_6)^2}{2a_7^2}\right)+\exp\left(-\frac{(\lambda-a_3+a_6)^2}{2a_7^2}\right)\right]$$

Here $\lambda$ is the wavelength in μm. The parameter $a_1$ represents the continuum intensity. $a_2$ and $a_5$ denote the amplitude of the central π component and the two side σ components, respectively. $a_3$ represents the wavelength offset of the central π component relative to the Mg I 12.32 μm line center. $a_4$ and $a_7$ correspond to the Gaussian standard deviations of the π and σ components, respectively. $a_6$ denotes the Zeeman splitting.

The above model parameters were estimated using a MCMC approach, which simultaneously sampled the posterior distributions of the seven free parameters ($a_1$–$a_7$). The magnetic field strength B was calculated using: $B = a_6/(4.67 \times 10^{-9} g\lambda^2)$, where g indicates the effective Landé factor and is 1 for the Mg I 12.32 μm line. The MCMC sampling provides the posterior probability distribution of $a_6$ (Extended Data Fig. 1), allowing its uncertainty to be quantified. The uncertainty in the magnetic field strength was subsequently obtained through standard error propagation from the posterior distribution of $a_6$. The resulting central 68% credible-interval half-width is 19.73 G for Extended Data Fig. 1, corresponding to the pixel 82 at 04:11:35 UT.

The temporal evolution of the mid-infrared continuum intensity and magnetic field strength for the nine pixels 76–84 during the flare, along with their fitted curves, is shown in Extended Data Fig. 2. Using the 9th-degree polynomial fitted curves, the continuum enhancement $E_I$, initial magnetic-field strength $B_0$, peak-to-peak magnetic-field decay ΔB, and the corresponding background-corrected decay $\Delta B_{fit\text{-}bg}$ are summarized in Extended Data Table 1.

To compare the flare-related magnetic-field decay during the impulsive phase with the background field variations at other times, we adopted Welch's t-test for its statistical significance. A second-order polynomial background fitting was performed to determine the baseline trend of the field strength for each pixel. The fitting was conducted over the time interval of 04:01:31-04:38:32 UT, with the flare-affected period of 04:09:20-04:18:19 UT masked out. We then subtracted the baseline trend and applied a one-side Welch test to the detrended evolutions of magnetic field. The resulting $p_{decay}$ value indicates how well the flare-related magnetic field decay can be distinguished from background variations. A smaller value means a more significant distinction. $p_{decay} < 0.05$ is used as a commonly adopted conventional criterion for statistical significance in the individual-pixel tests. For normalized continuum intensity, the enhancement was measured from the fitted intensity curve over the same analyzed interval. All relevant quantities, including the continuum intensity enhancement $E_I$, the initial magnetic-field strength $B_0$, the magnetic-field decay $\Delta B$, the baseline detrended magnetic-field decrease $\Delta B_{fit-bg}$, the background magnetic fields variation $\sigma_{bg}$, and the Welch-test $p_{decay}$ values, are summarized in the Extended Data Table 1.

The strongest statistical results are found in pixels 80-84 among the nine pixels, where $p_{decay}$ ranges from 1.7e-05 to 2.42e-04. Pixels 77-84 all have $p_{decay} < 0.05$, satisfying the conventional significance criterion for an individual one-sided test. The $p_{decay}$ is 0.061 for pixel 76, indicating marginal significance that does not meet the conventional $p < 0.05$ criterion. As illustrated in Fig. 1d, pixels 76-79 lie at greater distances from the magnetic PIL than pixels 80-84, with pixel 76 showing the maximum separation from the PIL. Such disparities in pixel-PIL separation suggest that the magnetic field topology could influence both the background field variations $\sigma_{bg}$ and the $p_{decay}$ values. Across individual pixels, $\Delta B$ varies between -53.89 G and -148.40 G, whereas $\Delta B_{fit-bg}$ ranges from -57.46 G to -138.97 G. The background field variations $\sigma_{bg}$ is larger than the fitting uncertainties derived from the MCMC approach. The discrepancy arises because our calculation of $\sigma_{bg}$ incorporates the additional contributions, i.e., the inherent magnetic field evolution during the flare, the variable residual stripe-like artifacts, transient flare spectral evolution over the FTIR scan window (30s) and the tracking errors (about 0.88”) of AIMS telescope.

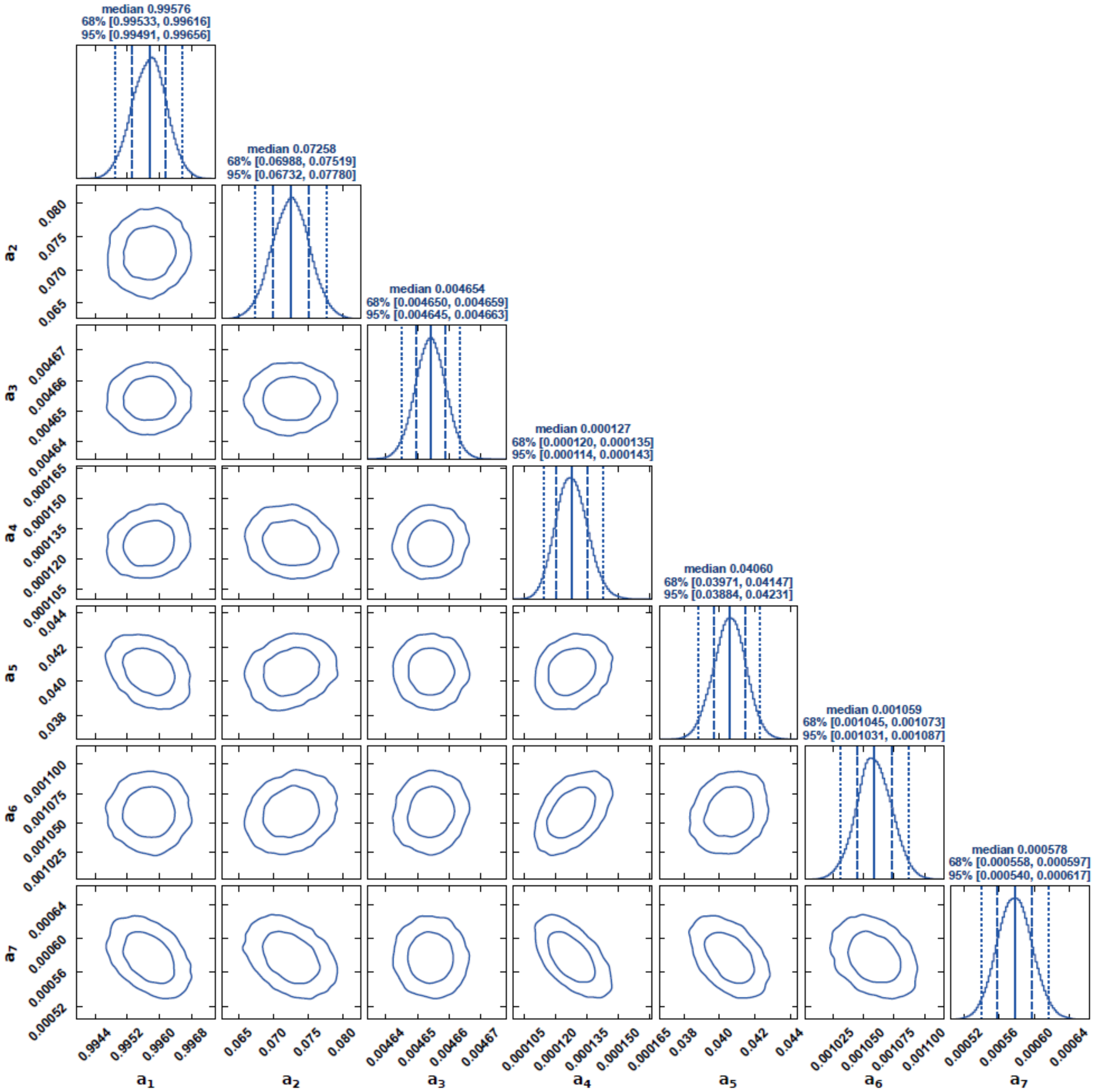

Extended Data Fig. 1 The seven fitted quantities are: ($a_1$), continuum intensity; ($a_2$), amplitude of the central π component; ($a_3$), line-centre offset relative to 12.31719764 μm; ($a_4$), Gaussian standard deviation of the π component; ($a_5$), common amplitude of the two σ components; ($a_6$), displacement of each σ peak from the π centre (the one-sided Zeeman splitting); and ($a_7$), common Gaussian standard deviation of the σ components. Each diagonal panel shows a one-dimensional marginal posterior. The solid line is the median, the dashed lines are the 16th and 84th percentiles, and the dotted lines are the 2.5th and 97.5th percentiles. Each title reports the median once and then the bounds of the central 68% and 95% credible intervals. The off-diagonal panels show joint marginal posteriors; the inner and outer contours enclose 68% and 95% of the posterior probability, respectively.

Extended Data Table 1. Statistical characterisation of the flare-associated magnetic field decay for pixels 76-84. $E_I$ denotes the peak-to-peak continuum enhancement derived from the fitted normalized-intensity curve over the time window 04:09:20-04:18:19 UT. $B_0$ and $\Delta B$ are the initial magnetic-field strength and the peak-to-peak magnetic-field decrease extracted from the fitted magnetic-field curve, respectively. $\Delta B_{fit-bg}$ refers to the peak-to-peak decrease obtained after subtracting the second-order background trend from the same fitted curve. Negative $\Delta B$ values indicate a reduction in magnetic-field strength. $\sigma_{bg}$ is the standard deviation of the background-corrected magnetic-field variations outside the flare-affected temporal interval. $p_{decay}$ indicates the one-sided Welch's t-test p-value, which quantifies the statistical significance of magnetic field decay during the flare against the background-magnetic field variability.

| Pixels | $E_I$ | B (G) | ΔB (G) | $\Delta B_{fit-bg}$ (G) | $\sigma_{bg}$ (Gauss) | $p_{decay}$ |
|---|---|---|---|---|---|---|
| 76 | 10.08% | 1395.87 | -148.40 | -138.97 | 107.51 | 0.061 |
| 77 | 12.32% | 1528.36 | -132.50 | -118.66 | 81.40 | 0.019 |
| 78 | 14.04% | 1593.92 | -119.12 | -112.12 | 80.14 | 0.008 |
| 79 | 15.46% | 1617.27 | -111.58 | -106.14 | 68.68 | 0.002 |
| 80 | 16.82% | 1635.48 | -107.55 | -104.16 | 56.84 | 1.14e-04 |
| 81 | 17.81% | 1635.67 | -117.88 | -118.28 | 59.50 | 1.62e-04 |
| 82 | 17.31% | 1644.79 | -111.13 | -116.22 | 64.11 | 2.42e-04 |
| 83 | 15.01% | 1662.27 | -83.16 | -88.19 | 43.15 | 1.70e-05 |
| 84 | 11.32% | 1665.21 | -53.89 | -57.46 | 36.42 | 6.89e-05 |

**Radiative-hydrodynamic simulations of flare spectrum**

We employ the radiative hydrodynamics code RADYN to calculate the atmospheric response to a typical flare driven by a beam of non-thermal electrons (refer to Model f10 [36] for details). The polarized Mg I line profiles are synthesized with the RH 1.5D code based on the RADYN output of the atmosphere plus a prescribed time-independent magnetic field. The response function of the splitting is calculated by perturbing the magnetic field at each height by 1%, resynthesizing the line profile, remeasuring the splitting, and dividing the resulted change by the imposed perturbation.

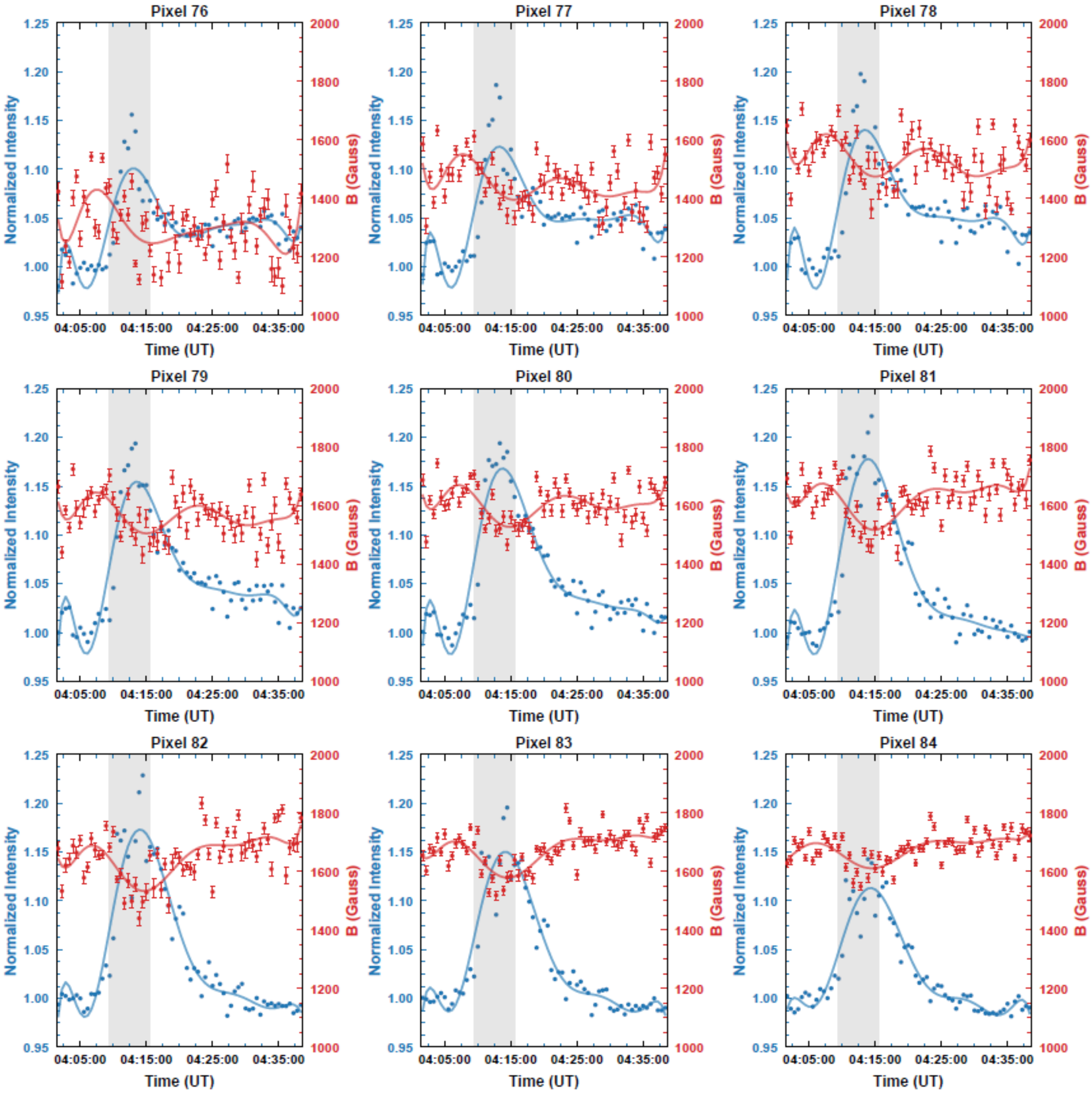

Extended Data Fig. 2 In each panel, the blue dots represent the temporal variation of the normalized mid-infrared continuum intensity,

and the blue line shows the 9th- polynomial fitting results. Red points show the MCMC-derived magnetic-field strength and the red curve is its 9th- polynomial fit. The red error bars are the formal spectral-fitting uncertainty propagated from the central 68% posterior interval of ($a_6$). The grey shading marks 04:09:20–04:15:31 UT, the interval over which the fitted magnetic field decreases.

## Inversion of the observed Mg I 12.32 μm flare spectra

To derive the temperature variations during the flare from the Mg I 12.32 μm spectra, we employed the parallel NLTE inversion code STiC[21], which uses the RH code[44] to synthesize Mg I 12.32 μm line profiles. The calculation of atomic populations in RH code assumes statistical equilibrium and a plane-parallel geometry. We note that STiC inversions are performed under the assumption of hydrostatic equilibrium, which may not be ideal in flares. Nevertheless, it remains the most optimal available approximation in the STiC code and has been applied to NLTE line inversions in multiple flare studies[25,45-46], representing a reasonable approximation.

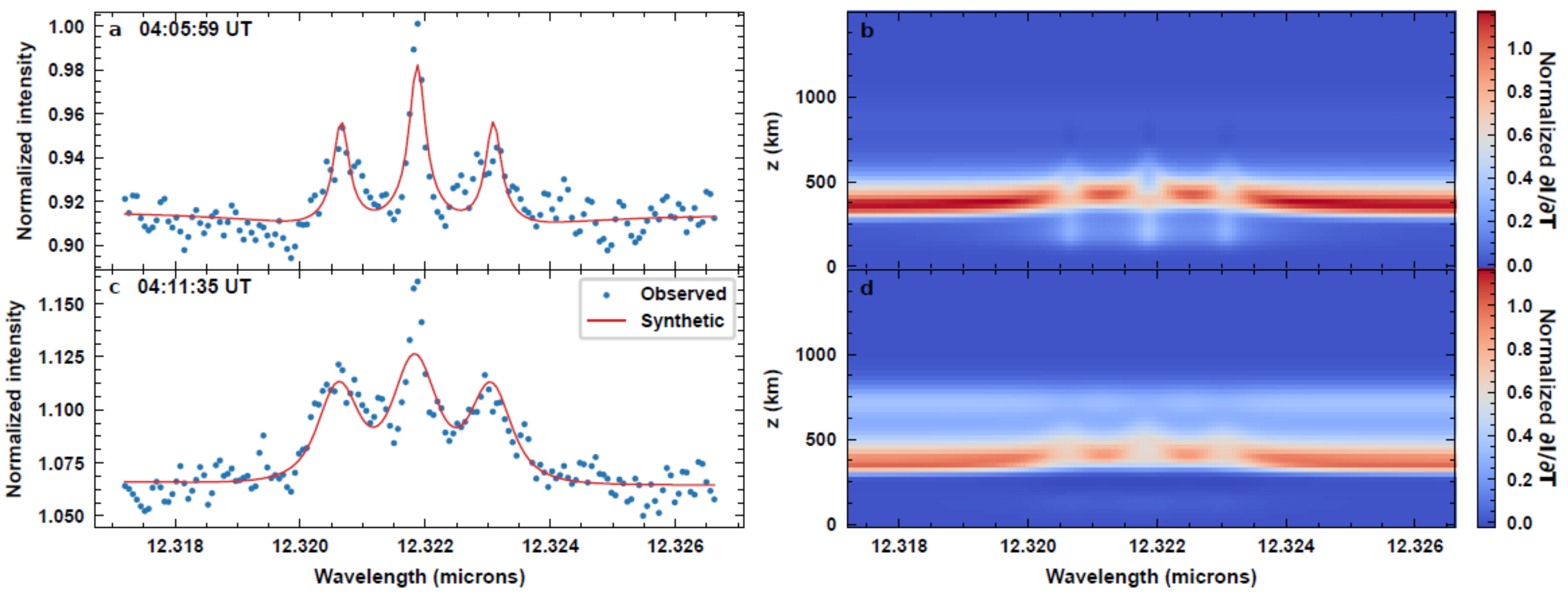

Extended Data Fig. 3 a and c: Observed (blue dots) and fitted results (red solid line) for 04:05:59 UT and 04:11:35 UT for pixel 82. b and d: Height distribution of the normalized temperature-intensity response function containing both continuum and the Mg I 12.32 μm line calculated with the STiC inverted atmospheric model.

## Data availability

ASO-S/HXT observations are available at http://aso-s.pmo.ac.cn/sodc/dataArchive.jsp. SDO data can be accessed via the Joint Science Operations Center at http://jsoc.stanford.edu/ajax/lookdata.html. GOES data can be downloaded from https://data.ngdc.noaa.gov/platforms/solar-space-observing-satellites/goes. The processed minimum dataset required to reproduce Figs. 1–4 and Extended Data Figs. 1–3, together with the source data underlying these figures are available in Zenodo at 10.5281/zenodo.22094379[47].

## Acknowledgements

We thank the Lenghu Astronomical Observatory for their long-term assistance and support in the construction, operation, and maintenance of AIMS. We thank the SDO, GOES, and ASO-S team for providing open access data. The AIMS data used in this research is observed by the assistance of Guoping Wang, Zongtai Tang, Huili Che, Xingming Bao, and Chengsen Wang.

## Funding Statement

The AIMS project is an instrument development project supported by the National Natural Science Foundation of China (NSFC) (grant No. 11427901). X.Y. Bai is funded by NSFC under Grant No. 12522306. This research is supported by the Strategic Priority Research Program of the Chinese Academy of Sciences (Grant No. XDB0560000), the National Key R&D Program of China (grant No. 2021YFA1600500) and the Specialized Research Fund for State Key Laboratories. J.Hong is funded by the European Union through the European Research Council (ERC) under the Horizon Europe program (MAGHEAT, grant agreement 101088184). The Institute for Solar Physics is supported by a grant for research infrastructures of national importance from the Swedish Research Council (registration number 2021-00169). W. Li acknowledges support from NSFC (grant No. 12373058).

## Author contributions

Y.Y.D. proposed the research idea. Y.Y.D. and X.Y.B. organized the research and paper writing, with X.Y.B.

also coordinating data analysis and processing. J.H. was responsible for simulations using the radiative hydrodynamics code RADYN. W.X.L. handled work related to the NLTE inversion code STiC. J.L.W., Y.Y.Y., K.F.J., Z.Y.H., and S.L.W. were responsible for data analysis. J.T.S. and Y.Y.Song provided interpretations of the observational results. Y.L.Shen, X.Y., X.M.Z., X.H., and B.Y.G. organized and implemented the observations. D.G.W., J.B.L., J.F.H., Z.W.F., W.D., L.Y.T., Y.B., J.W.H., L.W.Z., and S.B.X. are core members of the AIMS technology development team and contributed to data acquisition and calibration. All authors participated in discussions around the data analysis, results, and the article.